\documentclass[12pt,notcite, notref]{article}

\usepackage{graphics,amsmath,amssymb,amsthm,accents}
\usepackage{epic}

\def \h#1{\widehat{#1}}
\def \t#1{\widetilde{#1}}
\def \b#1{\overline{#1}}

\numberwithin{equation}{section}

\newcommand{\comment}[1]{}

\title{Boussinesq-like multi-component lattice equations\\ and
  multi-dimensional consistency
}

\author{Jarmo Hietarinta\footnote{E-mail: jarmo.hietarinta@utu.fi}
  \\Department of Physics and Astronomy\\ University of Turku,
  FIN-20014 Turku, Finland\footnote{permanent address}\\ and\\ Faculty
  of Mathematics, Kyushu University\\ 744 Motooka, Fukuoka 819-8581,
  Japan}

\date{\today}

\begin{document}

\maketitle

\begin{abstract}
We consider quasilinear, multi-variable, constant coefficient, lattice
equations defined on the edges of the elementary square of the
lattice, modeled after the lattice modified Boussinesq (lmBSQ)
equation, e.g., $\t y z=\t x-x$. These equations are classified into
three canonical forms and the consequences of their multidimensional
consistency (Consistency-Around-the-Cube, CAC) are derived. One of the
consequences is a restriction on form of the equation for the $z$
variable, which in turn implies further consistency conditions, that
are solved. As result we obtain a number of integrable multi-component
lattice equations, some generalizing lmBSQ.
\end{abstract}

\section{Introduction}
Recent progress in the theory of integrable difference equation has
been quite rapid, partly due to the effectiveness of the
Consistency-Around-the-Cube (CAC) property as an integrability
criterion \cite{CAC1,CAC2,CAC3}.  The work on CAC has so far been
mostly restricted to one-component equations, but some results do
exist for the three-component lattice Boussinesq (lBSQ) class of
equations \cite{NPCQ,N-DIGP,Wal,TN04,MK}. By this class we mean
equations relating variables defined on the corners of the elementary
square of the cartesian lattice, with some equations residing on the
edges and some on the square.  (In the following we will denote the
dependent variables by $x,y,z$ and the independent variables $n,m,k$,
furthermore shifts in the $m$-variable will be denoted by a tilde, in
the $n$-variable by a hat and in the $k$-variable by a bar, for
example $\b{\t x}\equiv x_{n+1,m,k+1}$.)

One example is the lBSQ (see equations (9) in \cite{TN04})
\begin{subequations}\label{lBSQ-l}
\begin{align}\label{lBSQ-l-q}
\t y&=x\t x-z,\quad
\h y=x\h x-z,\\
&\h{\t z}=x\h{\t x}-y+\frac{p^3-q^3}{\h x-\t x}.\label{lBSQ-l-z}
\end{align}
\end{subequations}
Equations \eqref{lBSQ-l-q} are defined on the sides, and the set can
be completed by suitable shifts, while \eqref{lBSQ-l-z} is defined on
the square.  Similarly the lattice modified BSQ/Schwarzian BSQ (see
equations (5.3.19-20) in \cite{Wal} and (4.7-8) in \cite{N-DIGP}) is
defined by
\begin{subequations}\label{mBSQ-l}
\begin{equation}\label{mBSQ-l-q}
\t y\, z=\t x-x,\quad \h y\, z=\h x-x,
\end{equation}
\begin{equation}\label{mBSQ-l-z}
\h{\t z}=\frac{z/y}{\t z-\h z}(\h z\t y p^3-\t z\h y q^3).
\end{equation}
\end{subequations}
(The second equations (\ref{lBSQ-l-z},\ref{mBSQ-l-z}) can be written
in other forms by replacing some shifted variables with others, using
the first equations.)

If one eliminates two of the variables and expresses the equation in
terms of the third one, the final equation turns out to be defined on
a $3\times 3$ stencil on the discrete 2D-plane. In fact, the lattice
modified and Schwarzian BSQ equations differ only in their
one-component $3\times3$ formulation: lmBSQ ((4.9) in \cite{N-DIGP})
is obtained after eliminating $x,z$, and lSBSQ ((4.1) in
\cite{N-DIGP}) after eliminating $y,z$.\footnote{From this point of
  view the edge equations are considered as Cole-Hopf-like
  transformations between the main equations, lmBSQ and lSBSQ.}
Further examples of this type are given in \cite{MK}.

From the point of view of the CAC approach we can take the quasilinear
equations as being given on all 12 edges of the consistency cube. In
the case of \eqref{mBSQ-l-q} they are
\begin{subequations}
\begin{align}
\t y\, z=\t x-x,\quad \h y\, z=\h x-x,\quad \b y\, z=\b x-x,\label{ed-1}\\
\h{\t y}\,\h z=\h{\t x}-\h x,\quad \b{\h y}\,\b z=\b{\h x}-\b x,
\quad \b{\t y}\, \t z=\b{\t x}-\t x,\label{ed-2}\\
\b{\t y}\,\b z=\b{\t x}-\b x,\quad \h{\t y}\,\t z=\h{\t x}-\t x,
\quad \b{\h y}\, \h z=\b{\h x}-\h x,\label{ed-3}\\
\b{\h{\t y}}\,\b{\h z}=\b{\h{\t x}}-\b{\h x},\quad 
\b{\h{\t y}}\,\b{\t z}=\b{\h{\t x}}-\b{\t x},\quad 
\b{\h{\t y}}\,\h{\t z}=\b{\h{\t x}}-\h{\t x}.\label{ed-4}
\end{align}
\end{subequations}
Equations \eqref{ed-1} should be considered as constraints on the
initial data: $x,y,\t y,\h y,\b y,z,\t z,\h z,\b z$ may be arbitrary,
but $\t x,\h x,\b x$ are restricted by \eqref{ed-1}. Equations
(\ref{ed-2},\ref{ed-3}) can next be used to determine the doubly shifted
$x,y$ quantities, resulting with the face equations 
\begin{equation}\label{mBSQ-xy}
\h{\t x}=\frac{\h x\,\t z-\t x\,\h z}{\t z-\h z},\quad
\h{\t y}=-\frac{\t x-\h x}{\t z-\h z},
\end{equation}
and the corresponding formulae for $\b {\t x},\, \b{ \h x},\, \b {\t
  y},\, \b {\h y}$. Then we can solve for $\b{\h{\t y}}\,\b{\h{\t x}}$
from \eqref{ed-4} and the 3D consistency implies, through the
third equation, a constraint on $z$:
\begin{equation}\label{mBSQ-zeq}
\frac{\h{\t z}(\t z-\h z)}{\t z\,\h z}+
\frac{\b{\h z}(\h z-\b z)}{\h z\,\b z}+
\frac{\t{\b z}(\b z-\t z)}{\b z\,\t z}=0.
\end{equation}
This can be separated and we find that
\begin{equation}\label{mBSQ-za}
\h{\t z}=\frac{\t z\,\h z }{\t z-\h z }(F_p-F_q),
\end{equation}
where the $F$'s are some free functions of the parameters and shifted
variables of the direction  indicated by the subscript, i.e.,
\begin{equation}\label{f-phi}
F_p=\phi(x,y,z,\t x,\t y,\t z,p),\quad
F_q=\phi(x,y,z,\h x,\h y,\h z,q),\quad
F_r=\phi(x,y,z,\b x,\b y,\b z,r).
\end{equation}
(Here some of the shifted variables can be eliminated by using \eqref{ed-1}.)

Up to this point we have only used the quasilinear equations and their
consequences. The 3D consistency of $z$-shifts, as derived from
\eqref{mBSQ-za} and the corresponding other shifts, implies further
requirements on the $\phi$, which in the case of lmBSQ/lSBSQ are solved
by (c.f., \eqref{mBSQ-l})
\begin{equation}\label{mBSQ-F}
\phi(x,y,z,\t x,\t y,\t z,p)=\frac{\t y\, z}{\t z \,y}p^3.
\end{equation}

In this paper we formalize and generalize the above point of view of
starting with quasilinear equations on the sides of the consistency
cube and then deriving multi-dimensionally consistent multi-component
equations. We start in Section 2 by classifying the quasilinear part
(under the constant coefficient assumption) and find that there are
only three essentially different cases to consider. In Section 3 we
derive for each case the consistency equations for $z$ and finally in
Section 4 solve these equations.

\section{The quasilinear part}
\subsection{Classification}
The following properties will be required of the quasilinear equations:
\begin{enumerate}
\item The equations are linear in both the shifted and the unshifted variables.
\item Precisely two shifted variables appear, we call them $x$ and $y$.
\item The third (non-shifted) variable $z$ appears in a nontrivial way.
%\end{enumerate}
%In this paper we further assume that \begin{enumerate}
\item The coefficients of the
quasilinear equations are all constants, or can be transformed to
constants by suitable ($n,m,k$-dependent) transformations.
\end{enumerate}
The second property is modeled after the existing lattice BSQ results,
while the fourth property is a technical restriction. It is possible
that other integrable models can be found if either or both of these
conditions are relaxed.

Clearly the most general equation satisfying the above requirements
has the form
\[
E:=(\t x,\t y,1)\begin{pmatrix}
a_{11} & a_{12} & a_{13} & b_{1}\\
a_{21} & a_{22} & a_{23} & b_{2}\\
c_{1} & c_{2} & c_{3} & d
\end{pmatrix}
\begin{pmatrix} x\\y\\z\\1 \end{pmatrix}=0.
\]

In order to find the canonical forms we have at our disposal
rotations in $x,y$, affine transformations in each coordinate, and
$z\mapsto o_1 x+o_2 y+o_3 z+t_z$. Furthermore we can apply rational
linear transformations of the type
\[
(x,y,z)\mapsto \left(\tfrac{P}{S},\tfrac{Q}{S},\tfrac{z+R}{S}\right)
\]
where $P,Q,R,S$ are linear functions in $x,y$.  Finally we may
sometimes use $n,m,k$ dependent transformations, provided that the
final result does not explicitly depend on $n,m,k$.

In Appendix A it is shown that all equations of the above type can be
transformed into one of the following three forms:
\begin{equation} 
{  \t x z=\t y+x,}\tag{A} \label{e-11}\\
\end{equation}
\begin{equation}
  \label{e-c5}
{  \t x\,x=\t y+z,}\tag{B}\\
\end{equation}
\begin{equation}
  \label{e-12fin}
{    \t y z=\t x-x.}\tag{C}
\end{equation}
The convention here is that the $x$-variable appears both with and
without a shift, $y$ only with a shift and $z$ only without. There is
a fourth case of this type, namely $\t x+x\t y=z$, but it can be
transformed to (A) using $(x,y,z)\mapsto (1/x,y/x,z/x)$.

\subsection{Reversal symmetry.}
The starting quasilinear equations have a reversal
symmetry. Take for example Case B, we have
\[
\t x x=\t y+z \,\stackrel{rev}{\longrightarrow}\,
{\underaccent{\tilde} x} x=\underaccent{\tilde}y+z\,
\stackrel{shift}{\longrightarrow}\, x\t x=y+\tilde z\,
\stackrel{z\leftrightarrow y}{\longrightarrow}\, \t x x=\t y +z.
\]
Thus the reversal must be accompanied with the exchange
$z\leftrightarrow y$. Here we cannot yet determine whether $x$ or the
spectral parameter should also change sign, but for the solutions
found later on it turns out that $x$ should change sign but spectral
parameters should not.

Similarly for Case C we have
\[
\t y z=\t x-x \,\stackrel{rev}{\longrightarrow}\,
{\underaccent{\tilde} y} z=\underaccent{\tilde}x-x\,
\stackrel{shift}{\longrightarrow}\, y\t z=x-\t x\,
\stackrel{y\leftrightarrow -z}{\longrightarrow}\, \t y z=\t x -x,
\]
Here, instead of $(x,y,z)\mapsto(x,-z,y)$ we could have other sign choices.

Case A is somewhat different, for first we have
\[
\t x z=\t y+x \,\stackrel{rev}{\longrightarrow}\,
{\underaccent{\tilde} x} z=\underaccent{\tilde}y+x\,
\stackrel{shift}{\longrightarrow}\, x\t z=y+\t x,
\]
after which we must make the slightly more complicated transformation
to get \eqref{e-11}, namely
\[
(x,y,z)\mapsto (1/x,-z/x,-y/x).
\]

\subsection{Evolution}
\begin{figure}[t]
%\begin{center}
\setlength{\unitlength}{0.0044in}
\begin{picture}(500,500)(-100,100)
\thinlines
\drawline(0,200)(400,200)
\drawline(0,400)(400,400)
\drawline(0,500)(400,500)
\drawline(50,550)(50,150)
\drawline(250,550)(250,150)
\drawline(350,550)(350,150)
\drawline(0,300)(400,300)
%\drawline(450,550)(450,150)
\drawline(150,550)(150,150)
\put(420,280){\makebox(0,0)[lb]{$n$}}
\drawline(400,300)(385,310)
\drawline(400,300)(385,290)
\put(100,550){\makebox(0,0)[lb]{$m$}}
\drawline(150,550)(160,535)
\drawline(150,550)(140,535)
\thicklines
\drawline(0,500)(50,500)
\drawline(50,500)(50,400)
\drawline(50,400)(150,400)
\drawline(150,400)(150,300)
\drawline(150,300)(250,300)
\drawline(250,300)(250,200)
\drawline(250,200)(350,200)
\drawline(350,200)(350,150)
\put(250,300){\circle*{12}}
\put(250,400){\circle*{12}}
\put(150,300){\circle*{12}}
\put(150,400){\circle*{12}}
\put(100,305){\makebox(0,0)[lb]{$W$}}
\put(100,405){\makebox(0,0)[lb]{$\h W$}}
\put(250,305){\makebox(0,0)[lb]{$\t W$}}
\put(250,405){\makebox(0,0)[lb]{$\h{\t W }$}}
\put(200,100){\makebox(0,0)[lb]{a)}}
\end{picture}\hskip 1cm
\begin{picture}(700,500)(-100,100)
\drawline(0,200)(400,200)
\drawline(0,400)(400,400)
\drawline(0,500)(400,500)
\drawline(50,550)(50,150)
\drawline(250,550)(250,150)
\drawline(350,550)(350,150)
\drawline(0,300)(400,300)
%\drawline(450,550)(450,150)
\drawline(150,550)(150,150)
\put(420,280){\makebox(0,0)[lb]{$n$}}
\drawline(400,300)(385,310)
\drawline(400,300)(385,290)
\put(100,550){\makebox(0,0)[lb]{$m$}}
\drawline(150,550)(160,535)
\drawline(150,550)(140,535)
\thicklines
\drawline(0,200)(50,200)
\drawline(50,200)(50,300)
\drawline(50,300)(150,300)
\drawline(150,300)(150,400)
\drawline(150,400)(250,400)
\drawline(250,400)(250,500)
\drawline(250,500)(350,500)
\drawline(350,500)(350,550)
\put(250,300){\circle*{12}}
\put(250,400){\circle*{12}}
\put(150,300){\circle*{12}}
\put(150,400){\circle*{12}}
\put(100,305){\makebox(0,0)[lb]{$W$}}
\put(100,405){\makebox(0,0)[lb]{$\h W$}}
\put(250,305){\makebox(0,0)[lb]{$\t W$}}
\put(255,405){\makebox(0,0)[lb]{$\h{\t W }$}}
\put(200,100){\makebox(0,0)[lb]{b)}}
\end{picture}
%\end{center}
\caption{Possible definitions of initial values and evolution. Here
  $W$ stands for the vector $(x,y,z)$. When equations are given on the
  links connecting lattice points some of them imply constraints on
  the initial values while some give part of the evolution.\label{F1}}
\end{figure}
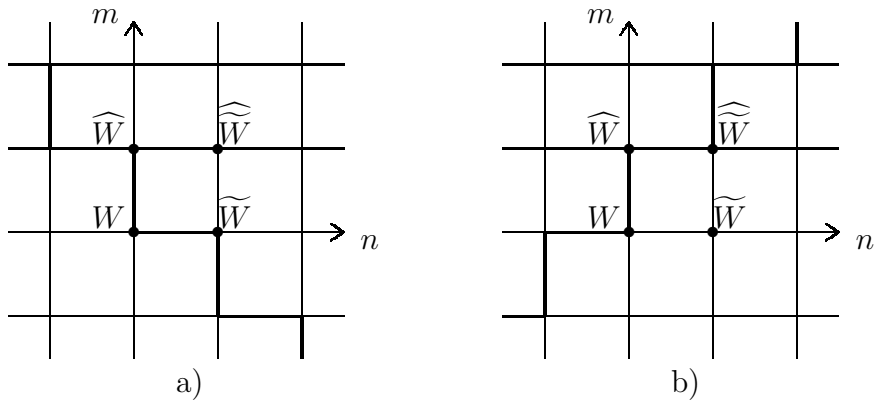

For an equation defined on the elementary square of the cartesian
lattice one usually considers initial values given on a staircase-like
configuration, see Figure \ref{F1}. We can, for example, have the
initial values given on a staircase extending from lower right to
upper left and with evolution towards upper right, as in Figure
\ref{F1}a). On the elementary square it means that the values with no
shift or one shift are given as initial values and the double shifted
quantities are to be derived from the equations.

In the present quasilinear case, however, we have four equations
defined on the edges of the elementary square and two of the equations
only involve the initial values; these equations are therefore to be
considered as constraints on the initial data. Thus we can take, for
example, $x,y,z,\t y,\t z, \h y,\h z$ as given and then determine $\t
x,\h x$ using two of the four edge equations, after which the
remaining two edge equations give $\h{\t x},\h{\t y}$, for example as
in \eqref{mBSQ-xy}. If, on the other hand, we were to consider an
initial staircase extending from lower left to upper right and
evolution towards lower right, as in Figure \ref{F1}b), we could take
$\h x,\h y,\h z, y, z, \h{\t y},\h{\t z}$ as given, $x,\h{\t x}$
constrained by two edge equations, and then get from the remaining two
edge equations two relations involving $\t x,\t y,\t z$ as part of the
evolution towards lower right. However, in the present case the edge
equations are not symmetric, since they involve $y$ and $z$ in a
different way, and therefore the equations giving evolution towards
lower right and those evolving towards upper right have different
structure.  In the following we will only consider the evolution of
Figure \ref{F1}a) and its completion to the consistency cube.

So far the missing equation in the evolution is the one for $\h{\t z}$,
its form will be determined in the following from considerations
of multidimensional consistency.

\section{CAC}
We will now derive the consequences of multidimensional consistency
for the basis equations.

\subsection*{Case A}
Writing out the starting quasilinear equations \eqref{e-11} we
have
\begin{equation}
  \label{e-c1.1all}
  \t x\, z=\t y+x,\quad
  \h x\, z=\h y+x,\quad
  \b x\, z=\b y+x.
\end{equation}
Now taking various shifts of these equations and then solving for the
doubly shifted quantities we get \comment{see {\tt (ajo.f11)}}
\begin{equation}
  \label{e-c1.1-pq}
  \h{\t x}=\frac{\t x-\h x}{\t z-\h z},\quad
  \h{\t y}=\frac{\t x\,\h z-\h x\,\t z}{\t z-\h z},
\end{equation}
and similar equations for the bar-tilde and bar-hat shifts. There are
many other ways to write these equations, by inserting suitable
combinations of the edge equation. For example, a reversal symmetric
form is obtained with
\begin{equation}
  \label{e-c1.1-pq-symm}
 \frac1x\,{ \h{\t x}}=\frac{(\t x-\h x)^2}{(\t z-\h z)(\h x\t y-\t x\h y)},\quad
 \frac{z}{x}\, \h{\t y}=\frac{(\t x\,\h z-\h x\,\t z)(\t y-\h y)}
{(\t z-\h z)(\h x\t y-\t x\h y)}.
\end{equation}

The requirement of 3D consistency on $x$ is derived from
\[
(\h{\t x})\,\bar{}=(\bar{\t x})\,\h{}=(\h{\bar x})\,\t{},
\]
This and the corresponding equations for $y$ lead to one single
condition \comment{(see {\tt ajo.f11})}\footnote{Actually the
  requirement factorizes, with the other factor being $\t x(\h z-\b
  z)+\h x(\b z-\t z)+\b x(\t z-\h z)$, but since this is a first order
  condition we will not use it. The same happens in the other cases.
  \comment{laske kuitenkin mihin se vie eri tapauksissa!}}
\begin{equation}
  \label{e-c1.1-cac}
   \h{\t z}(\t z-\h z)
+  \b{\h z}(\h z-\b z)
+  \t{\b z}(\b z-\t z)=0.
\end{equation}
Each of the three terms involve two of the three possible shifts and
we can therefore separate the variables with the solution
\begin{equation}
  \label{e-c1.1-zth}
  \h{\t z}= \frac1{\t z-\h z}(F_p-F_q),
\end{equation}
and similarly (cyclically) for the other shifts. Here $F_p,F_q,F_r$
are some (rational) functions associated with the $m,n,k$ directions
(i.e., tilde, hat and bar shifts), respectively, as described in
\eqref{f-phi}. In practice we will eliminate the possible $\t y,\h
y,\b y$ dependency using \eqref{e-c1.1all}.  The dependency of $\phi$ on
the shifted quantities must be nontrivial so that for example $\h{\t
  z}\neq\b{\t z}$.  In particular $F_p=\beta(x,y,z)\t z$ is not
acceptable.

Finally the requirement of multidimensional consistency of
\eqref{e-c1.1-zth} implies that the function $F$ just introduced must
satisfy \comment{ see {\tt f-aa} }
\begin{equation}
%\boxed{  
\label{e-c1.1-zcac}
  \frac{\h F_p-\h F_r}{\t z-\b z}
= \frac{\b F_q-\b F_p}{\h z-\t z}
= \frac{\t F_r-\t F_q}{\b z-\h z},
%}
\end{equation}
where $\h F_p=\phi(\h x,\h y,\h z,\h {\t x},\h {\t z},p)$, etc.
This is the main equation to be solved in the next chapter. Note again
that the spurious solution $F_p=\beta(x,y,z)\t z$ is not acceptable.
In solving this equation we should use the previous equations
\eqref{e-c1.1-pq},\eqref{e-c1.1-zth} to eliminate all doubly shifted
quantities, but we also have to eliminate shifted $y$'s using
\eqref{e-c1.1all}.

\subsection*{Case B}
The equations are
\begin{equation}
  \label{eq:B-base}
\t x\,x=\t y+z,\quad \h x\,x=\h y+z,\quad \b x\,x=\b y+z.
\end{equation}
After computing the various shifts of these equations we use them
again to eliminate shifted $y$'s from the results. The doubly shifted
variables then have the form
\begin{equation}
  \label{e-c5-pq}
  \h{\t x}=\frac{\t z-\h z}{\t x-\h x},\quad
  \h{\t y}=\frac{\h x\,\t z-\t x\,\h z}{\t x-\h x}.
\end{equation}
These equations are not symmetric under the reversal, but it is easy
to add linear combinations of \eqref{eq:B-base} to get
reversal symmetric forms, for example of the form
\begin{equation}\label{e-c5-pq-symm}
(\h{\t x}-x)(\t x-\h x)=(\t z-\t y)-(\h z-\h y),\quad
(\h{\t y}-z)(\t x-\h x)=(\t z-\t y)\h x-(\h z-\h y)\t x.
\end{equation}
In either case the consistency condition becomes \comment{f-cc}
\begin{equation}
  \label{e-c5-cac}
   \h{\t z}\,(\t x-\h x)
+  \b{\h z}\,(\h x-\b x)
+  \t{\b z}\,(\b x-\t x)=0,
\end{equation}
which is resolved by
\begin{equation}
  \label{e-c5-zpq}
  \h{\t z} = \frac1{\t x-\h x}(F_p-F_q).
\end{equation}
The 3D-consistency for $z$ then leads to \comment{using {\tt f-cc} }
\begin{equation}
  \label{e-c5-zcac}
%\boxed{
  \frac{\h F_p-\h F_r}{\t x-\b x}
= \frac{\b F_q-\b F_p}{\h x-\t x}
= \frac{\t F_r-\t F_q}{\b x-\h x}
%}
\end{equation}
This is similar to the result \eqref{e-c1.1-zcac} for Case A, but note
the different denominators, $\t x-\h x$ vs.~$\t z-\h z$.

\subsection*{Case C}
The starting equations are now given by
\begin{equation}
  \label{e-c1.2all}
  \t y\, z=\t x-x,\quad
  \h y\, z=\h x-x,\quad
  \b y\, z=\b x-x.\quad
\end{equation}
Following the procedure given above we first get \comment{using {\tt f-bb},
see also older {\tt ajo.f-12} }
\begin{equation}
  \label{e-c1.2-pq}
  \h{\t x}=\frac{\h x\,\t z-\t x\,\h z}{\t z-\h z},
\quad
  \h{\t y}=-\frac{\t x-\h x}{\t z-\h z}=-z\frac{\t y-\h y}{\t z-\h z}.
\end{equation}
Again it is easy to generate the  reversal symmetric forms, for example,
\begin{equation}
  \label{e-c1.2-pq-symm}
\h{\t x}(\t z-\h z)+x(\t y-\h y)=(\t y+\t z)\h x-(\h y+\h z)\t x,\quad
\h{\t y}(\t z-\h z)+z(\t y-\h y)=2(\t x-\h x).
\end{equation}
or
\begin{equation}
\h{\t x}=x-\frac{(\t x-\h x)(\t y\h z-\h y\t z)}{(\t y-\h y)(\t z-\h z)},\quad
\h{\t y}=z-\frac{(\t x-\h x)(\t y-\h y+\t z-\h z)}{(\t y-\h y)(\t z-\h z)}.
\end{equation}
From any form we get the consistency condition
\begin{equation}
  \label{e-c1.2-cac}
   \h{\t z}(\t z-\h z)\bar z
+  \b{\h z}(\h z-\b z)\t z
+  \t{\b z}(\b z-\t z)\h z=0.
\end{equation}
The form of $\h{\t z}$ must then be
\begin{equation}
  \label{e-c1.2-zth}
  \h{\t z}= \frac{\t z\h z}{\t z-\h z}(F_p-F_q)
\end{equation}
and its 3D consistency implies
\begin{equation}
%\boxed{  
\label{e-c1.2-zcac}
  \frac{\h F_p-\h F_r}{F_p-F_r}
= \frac{\b F_q-\b F_p}{F_q-F_p}
= \frac{\t F_r-\t F_q}{F_r-F_q}.
%}
\end{equation}
This is essentially different from Case A \eqref{e-c1.1-zcac} or Case
B \eqref{e-c5-zcac}.

\section{Solutions}
We will now proceed to solve the final CAC equations. They are all of
the form
\begin{equation}\label{e-ffeq}
%  \frac{F(p,q)-F(r,q)}{G(p)-G(r)} = \frac{F(q,r)-F(p,r)}{G(q)-G(p)} =
%  \frac{F(r,p)-F(q,p)}{G(r)-G(q)},
%\end{equation}
%\begin{equation}
%\label{e-c1.2-zcac}
  \frac{\h F_p-\h F_r}{G_p-G_r} = \frac{\b F_q-\b F_p}{G_q-G_p} =
  \frac{\t F_r-\t F_q}{G_r-G_q}.
\end{equation}
where the subscript indicates on which variables the function depends
on, c.f., \eqref{f-phi}.  The meaning of $G$ depends on the particular
case. In the shifted functions, for example $\h F_p=\phi(\h x,\h y,\h
z,\h {\t x},\h {\t z},p)$, we should use the previous formulas to
express $\h{\t x}$ and $\h{\t z}$ in terms of the once-shifted
quantities, and thus we should think $\h F_p$ as a function of
$x,y,z,\t x,\h x,\t z,\h z$. 

Equation \eqref{e-ffeq} contains functions of at most two of the three
sets of shifted variables appearing in the equation. Concentrating on
$p,q$ sets, say, (and fixing the values of $r,\b x,\b y,\b z$) we find
easily that $\h F_p$ must have the form \comment{see {\tt ffeq}}
\begin{equation}
  \label{e-ff-form}
  \h F_p=(G_p-C)\,\frac{A_p-A_q}{G_p-G_q}+B_q,
\end{equation}
where $A,B$ are some functions of the unshifted variables, and of the shifted
variables and parameters indicated by the subscript, while $C$ may
only depend on the unshifted variables. Using the same form for the
other $F$'s as well solves the equations \eqref{e-ffeq}.
Thus the final problem is to solve \eqref{e-ff-form}.

\subsection*{Case A}
Using (\ref{f-phi},\ref{e-c1.1-pq},\ref{e-c1.1-zth}) the explicit form of
\eqref{e-ff-form} is
\begin{equation}
  \label{eq:ff-exp-A}
  \phi\left(\h x,\h y,\h z,\frac{\t x-\h x}{\t z-\h z}, \frac{F_p-F_q}{\t z-\h
    z},p\right)=(\t z-C)\frac{A(\t x,\t z,p)-A(\h x,\h z,q)}{\t z-\h
  z}+B(\h x,\h z,q)
\end{equation}
where the dependence on unshifted variables is hidden. Let us first
consider the generic case in which $A(\t x,\t z)$ is more complicated
than $c\t z$. Then as a function of $\t x,\t z$ the RHS has a simple
pole at $\t z=\h z$.  Observing the form of the LHS we see that if
$F_p-F_q\not\propto \t x-\h x$ (i.e., $\h{\t z}\not \propto\h{\t x}$)
$\phi$ must be linear in the fourth and fifth variables, otherwise it
would also have other poles. Thus we find that
\begin{equation}\label{f-11}
F_p=\alpha_p\t x+\beta_p\t z+\gamma_p,
\end{equation}
where the functions $\alpha_p,\beta_p,\gamma_p$ may depend on
$x,y,z,p$ (but not on shifted quantities), and similarly for
$F_q,F_r$. This even includes the special cases $A=c\t z$ and
$F_p-F_q\propto \t x-\h x$.

Equation \eqref{eq:ff-exp-A} can now be written in the
form
\begin{align}
  \h \alpha_p(\t x-\h x)
+ \h \beta_p(\alpha_p\t x+&\beta_p\t z+\gamma_p-\alpha_q\h
  x-\beta_q\h z-\gamma_q)+ (\t z-\h z)\h \gamma_p\nonumber \\ =
&(\t z-C)(A(\t x,\t z,p)-A(\h x,\h z,q))+B(\h x,\h z,q) (\t z-\h z).
\label{a-eq-1}
\end{align}
Concentrating on the tilde-shifted quantities we find that
\begin{equation}
  \label{e-aa-11}
  A(\t x,\t z,p)=\frac{c_1(p)\t x+c_2(p)\t z+c_3(p)}{\t z-C},
\end{equation}
where the functions $c_i(p)$ may depend on $x,y,z$, and similarly for
the hat-shifted quantities. When these are used in \eqref{a-eq-1} the
coefficient of $\t x$ factorizes as $(\h z-C)[\h \alpha_p+\alpha_p\h
  \beta_p-c_1(p)]$.  In explicit form the equation to solve is
\begin{equation}
  \label{e-maineq-11}
  \alpha(\h x,\h x z-x,\h z,p)+\alpha(x,y,z,p)\beta(\h x,\h x z-x,\h
  z,p)
  =c_1(x,y,z,p),
\end{equation}
where we have already eliminated $\h y$ using \eqref{e-c1.1all}.
Equation \eqref{e-maineq-11} is solved in Appendix \ref{sol11}. Once
the solution for $\alpha,\,\beta$ is known it is relatively
straightforward to solve the remaining equations.

There are three different cases:

\paragraph{Case A.1:} For $\alpha=0$ the remaining equations imply $\beta=b_0+z,\,
\gamma=\gamma(p)$, leading to the solution  \comment{ {\tt ajo.fth-aa.1}}
\begin{equation}\label{sol-A-1}
%\boxed{    
\h{\t z}=b_0+z+\frac{p-q}{\t z-\h z}.
%}
\tag{A-1}
\end{equation}
where we have redefined the spectral parameters $\gamma(p)\to p$ etc.
Here $b_0$ can be eliminated by the transformation $z_{nm}\mapsto
z_{nm}+\tfrac12(n+m+k)b_0$, but in order to keep intact the form
\eqref{e-c1.1all} this needs to be accompanied with $y\mapsto
y+b_0\tfrac12(n+m+k)x$. 

Equation \eqref{sol-A-1} depends only on $z$ so the system is
triangular with $z$ driving the system, $x$ determined next from
\eqref{e-c1.1-pq} and $y$ last from \eqref{e-c1.1all}.  The
$z$-equation is that of the discrete potential KdV: $( \h{\t z}-z)(\t
z-\h z)=p-q.$

\paragraph{Case A.2:} For $\alpha=\alpha(p)\neq 0,\,\beta=\beta(p)$
the remaining equations imply that $\alpha,\beta$ are in fact
constants and $\gamma$ does not depend on lattice parameter. This
leads to a trivial solution
\begin{equation*}
  \h{\t z}=b_0+a_0\h{\t x},
  \text{ i.e., } 
  z=b_0+a_0 x,
\end{equation*}
which in turn implies that $\h{\t x}$ is constant.

\paragraph{Case A.3:} For the third case, starting with
$\alpha(x,y,z,p)=a(p)/x,\,\beta(x,y,z,p)=y/x+b(p)$, the conditions
from the remaining equations imply that $b(p)$ is a constant and that
$\gamma$ does not depend on lattice parameter. This leads to a genuine
coupled system with $\h{\t z}$ given by \comment{{\tt ajo.fth-aa.3}}
\begin{equation}\label{sol-A-3}
%\boxed
{\h{\t z}-y/x=b_0+\frac1x\frac{p\,\t x-q\,\h x}{\t z-\h z}.}
\tag{A-2}
\end{equation}
The parameter $b_0$ can be eliminated with $z_{nm}\mapsto
z_{nm}+b_0\tfrac13(n+m+k),\, y\mapsto y+b_0\tfrac13(n+m+k)x$, and we
have redefined $a(p)\to p$. This equation is symmetric under the
reversal if we replace the divisor $x$ in the last term using $x=(\t
x\h y-\h x\t y)/(\t x-\h x)$:
\begin{equation}\label{sol-A-3r}
%\boxed
\h{\t z}-y/x=\frac{(p\,\t x-q\,\h x)(\t x-\h x)}
{(\t z-\h z)(\t x\h y-\h x\t y)}.
\tag{A-2r}
\end{equation}
We recall that in this case the accompanying equations are
\eqref{e-c1.1all} together with \eqref{e-c1.1-pq} or
\eqref{e-c1.1-pq-symm}. 
Furthermore, the triply shifted quantities are
\begin{subequations}
\begin{eqnarray}
\b{\h{\t x}}&=&x\,\frac{
\t x(\h z-\b z)+
\h x(\b z-\t z)+
\b x(\t z-\h z)}{
p\t x(\h z-\b z)+
q\h x(\b z-\t z)+
r\b x(\t z-\h z)},\\
\b{\h{\t y}}&=&\frac{y}x\, \b{\h{\t x}}+ \frac{
p\t x(\h x-\b x)+
q\h x(\b x-\t x)+
r\b x(\t x-\h x)}{
p\t x(\h z-\b z)+
q\h x(\b z-\t z)+
r\b x(\t z-\h z)},\\
\b{\h{\t z}}&=&z-x\frac{
p(\h z-\b z)+
q(\b z-\t z)+
r(\t z-\h z)}{
p\t x(\h z-\b z)+
q\h x(\b z-\t z)+
r\b x(\t z-\h z)}.
\end{eqnarray}
\end{subequations}

\subsection*{Case B}
This case is similar to Case A, differing in the exchange of
$x\leftrightarrow z$ in some parts of the formulae. For example the equation
corresponding to \eqref{eq:ff-exp-A} now reads \comment{ {\tt ajo.fth5}}
\begin{equation}
  \label{eq:ff-exp-B}
  \phi\left(\h x,\h y,\h z,\frac{\t z-\h z}{\t x-\h x}, \frac{F_p-F_q}{\t x-\h
    x},p\right)=(\t x-C)\frac{A(\t x,\t z,p)-A(\h x,\h z,q)}{\t x-\h
  x}+B(\h x,\h z,q),
\end{equation}
and as before we conclude that $F_p$ must have the form
\eqref{f-11}. When this is used in \eqref{eq:ff-exp-B} we get,
corresponding to \eqref{a-eq-1}
\begin{align}
  \h \alpha_p(\t z-\h z)
+ \h \beta_p(\alpha_p\t x+&\beta_p\t z+\gamma_p-\alpha_q\h
  x-\beta_q\h z-\gamma_q)+ (\t x-\h x)\h \gamma_p\nonumber \\ =
&(\t x-C)(A(\t x,\t z,p)-A(\h x,\h z,q))+B(\h x,\h z,q) (\t x-\h x),
\label{b-eq-1}
\end{align}
and then
\begin{equation}
  \label{e-aa-11b}
  A(\t x,\t z,p)=\frac{c_1(p)\t x+c_2(p)\t z+c_3(p)}{\t x-C}.
\end{equation}
The coefficient of $\t z$ now yields
\begin{equation}
  \label{e-maineq-5}
  \alpha(\h x,\h x x-z,\h z,p)+\beta(x,y,z,p)\beta(\h x,\h x x-z,\h
  z,p)=c_2(x,y,z,p).
\end{equation}
This differs from \eqref{e-maineq-5} mainly by being quadratic in $\beta$.
It is solved in Appendix \ref{sol5} with the result \eqref{s5-res}.

When the solution \eqref{s5-res} is substituted into \eqref{b-eq-1}
the subsequent equations quickly imply\comment{{\tt ajo.fth5}} that
the parameters $a_0,b_0,b_1$ are in fact constants. There are two
solutions: 

\paragraph{Case B.1:} The first solution is obtained if $\gamma=0,\,b_1\neq 0$:
\begin{equation}\label{s-B-1}
  \h{\t z}=a_0-b_0b_1\,x-b_1^2\, y+(b_0+b_1\,x)\,\h{\t x}
\end{equation}
Recall that the other equations that go with the above are
\eqref{eq:B-base} and \eqref{e-c5-pq} or \eqref{e-c5-pq-symm}.

Equation \eqref{s-B-1} can be simplified by further transformations
which nevertheless must keep the equations \eqref{e-c5-pq} intact.
We find that there are two canonical forms, the first of which is
\comment{{\tt tra-bb1}}
\begin{equation}\label{s-B-1acan}
\h{\t z}=x\,\h{\t x}-y\tag{B-1.1}.
\end{equation}
If $b_1^2\neq 1$ we can use the linear transformation
\[
(x,y,z)\mapsto (x-\tfrac{b_0}{b_1-1},
y-x\tfrac{b_0}{b_1-1}+\tfrac{a_0(b_1-1)+b_0^2b_1}{(b_1+1)(b_1-1)^2},
z-x\tfrac{b_0}{b_1-1}+\tfrac{-a_0(b_1-1)+b_0^2}{(b_1+1)(b_1-1)^2}),
\]
and follow it with the scaling
$(x,y,z)\mapsto(s^{2(n+m+k)+1}x,s^{4(n+m+k)}y, s^{4(n+m+k)+4}y)$,
where $s^6=b_1$, to transform \eqref{s-B-1} into the form
\eqref{s-B-1acan}. If $b_1=-1$ we use instead
\[
(x,y,z)\mapsto (-x+\tfrac12{b_0},
y-x\tfrac12{b_0}+(\tfrac13a_0+\tfrac16b_0^2)(1-n-m-k),
z-x\tfrac12{b_0}+(\tfrac13a_0+\tfrac16b_0^2)(n+m+k)+\tfrac14 b_0^2),
\]
and scaling. If $b_1=1,b_0=0$ we get \eqref{s-B-1acan} directly.

If $b_1=1,b_0\neq 0$ we cannot eliminate $b_0$, only scale it, but we
can change the value of $a_0$ with $n,m,k$ dependent translation.
This yields the second form
\begin{equation}\label{s-B-1bcan}
\h{\t z}=(x+1)\,(\h{\t x}-1)-y\tag{B-1.2}.
\end{equation}

Both of the forms \eqref{s-B-1acan} and \eqref{s-B-1bcan} are
manifestly reversal symmetric, but the second case requires that $x$
changes sign under reflection.

The triply shifted quantities reveal that the system \eqref{s-B-1acan}
is in fact periodic on the diagonal of the cube:
\begin{equation}
  \label{eq:cac-B-1a}
  \b{\h{\t x}}=x,\quad \b{\h{\t y}}=y,\quad   \b{\h{\t z}}=z.
\end{equation}
In the other case \eqref{s-B-1bcan} we have a similar result
\begin{equation}
  \label{eq:cac-B-1b}
  \b{\h{\t x}}=x+1,\quad \b{\h{\t y}}=y+x+1,\quad   \b{\h{\t z}}=z+x.
\end{equation}

\paragraph{Case B.2:} The other solution is obtained for $\gamma=p$, $b_1=1$:
\begin{equation}\label{s-B-2}
\h{\t z}=a_0-b_0(x-\h{\t x})-y+x\,\h{\t x}+\frac{p-q}{\t x-\h x},\tag{B-2}
\end{equation}
together with \eqref{e-c5-pq}. Here the parameter $a_0$ can be
transformed away by an $n,m,k$ dependent translation, similar to the
ones used before.  For $b_0=0$ we recover lBSQ \eqref{lBSQ-l}. This
result is reversal symmetric, but if $x$ changes sign, as in the
previous results, then $p,q$ should not change sign.

The triply shifted quantities are \comment{{\tt cac-bb2}}
\begin{subequations}
\begin{eqnarray}
\b{\h{\t x}}&=&b_0+x+\frac{
(q-r)\t x+
(r-p)\h x+
(p-q)\b x}{
\t x(\h z-\b z)+
\h x(\b z-\t z)+
\b x(\t z-\h z)},\\
\b{\h{\t y}}&=&b_0x+y+\frac{
(q-r)\t z+
(r-p)\h z+
(p-q)\b z}{
\t x(\h z-\b z)+
\h x(\b z-\t z)+
\b x(\t z-\h z)},\\
\b{\h{\t z}}&=&z+b_0\b{\h{\t x}}-\frac{
(q-r)\h x\b x+
(r-p)\b x\t x+
(p-q)\t x\h x}{
\t x(\h z-\b z)+
\h x(\b z-\t z)+
\b x(\t z-\h z)}.
\end{eqnarray}
\end{subequations}

\subsection*{Case C}
For this case we have $G(p)=F_p$ in \eqref{e-ff-form}, i.e.,
\begin{equation}
  \label{eq:ff-exp-C}
  \phi\left(\h x,\h y,\h z,-z\frac{\t y-\h y}{\t z-\h z}, 
    \frac{\t z\h z(F_p-F_q)}{\t z-\h z} 
    ,p\right)=(F_p-C)\frac{A(\t y,\t z,p)-A(\h y,\h z,q)}{F_p-F_q}
  +B(\h y,\h z,q),
\end{equation}
It is now more convenient to eliminate shifted $x$-variables, since
they appear linearly, and therefore we use $\phi(x,y,z,\t y,\t z)$. In
the generic case the pole structure now implies
\begin{equation}\label{fp-cc}
F_p=[\alpha_p\t y+\beta_p\t z+\gamma_p]/\t z,
\end{equation}
and then \eqref{eq:ff-exp-C} becomes
\begin{align}
z(\h y-\t y)\h\alpha_p+
\h\beta_p(\alpha_p\t y\h z-&\alpha_q\h y\t z+
(\beta_p-\beta_q)\h z\t z+
\gamma_p\h z-\gamma_q\t z)
+(\t z-\h z)\h {\gamma_p}\nonumber\\
=\h z(\alpha_p& \t y+\beta_p\t z+\gamma_p-C\t z)[A(\t y,\t z,p)-A(\h
y,\h z,q)]+\nonumber\\
&B(\h y,\h z,q)[\alpha_p\t y\h z-\alpha_q\h y\t z+
(\beta_p-\beta_q)\h z\t z+
\gamma_p\h z-\gamma_q\t z].
\end{align}
Since the dependence on tilde-shifted quantities is still linear,
except for the $A$ term, we conclude
\begin{equation}
  \label{e-aa-11c}
  A(\t y,\t z,p)=\frac{c_1(p)\t y+c_2(p)\t z+c_3(p)}
{\alpha_p\t y+\beta_p\t z+\gamma_p-C\t z}.
\end{equation}
When this is substituted into \eqref{eq:ff-exp-C} the result is linear
in $\t y,\t z$ and a suitable linear combination of their
coefficients yields the equation
\begin{align}
  \alpha(\h y z+x,\h y,\h z,p)\,(z/\h z)\,[\alpha(x,y,z,p)\h
  y+\gamma(x,y,z,p)]
  -\gamma(\h y z+x,\h y,\h z,p)\,\alpha(x,y,&z,p)=\nonumber\\
  \alpha(x,y,z,p) c_3(x,y,z,p)-\gamma(x,y,z,p)c_1(x,y,z,p).&
  \label{eq:cc-eq}
\end{align}
This equation is solved in Appendix \ref{sol-cc} with four solutions,
$\alpha\equiv 0$, \eqref{eq:cc-r1}, \eqref{eq:cc-r2}, and \eqref{eq:cc-r4}.
The other equations for each case are \eqref{e-c1.2all} and
\eqref{e-c1.2-pq} or \eqref{e-c1.2-pq-symm}.

\paragraph{Case C.1:} $\alpha\equiv 0$. First assume that also
$\gamma\equiv 0$,  then the  remaining equations can be easily solved
with the result
\begin{equation}
  \label{eq:cc-r11}
  \h{\t z}=\frac{\t z\h z(p-q)}{\t z-\h z}.\tag{C-1.1}
\end{equation}
Next if $\gamma\not\equiv 0$ the analysis can be divided into sub-cases
according to whether or not $\partial_z\gamma=0$. In the former case
only trivial result follow ($\h{\t z}\propto z$), in the latter case
one gets the solution
\begin{equation}
  \label{eq:cc-r12}
  \h{\t z}=\frac{z(p\h z-q\t z)}{\t z-\h z}.\tag{C-1.2}
\end{equation}
Both of these results depend only on $z$ and its shifts so they form a
triangular system, as in Case A-1. In the case of \eqref{eq:cc-r12}
the equation can be scaled to lattice mKdV.

\paragraph{Case C.2:}
$\gamma(x,y,z,p)=\alpha(x,y,z,p)(x+g(p))/z$. Concentrating next on the
part multiplying $\t y$ and absorbing all $q$ dependency into $B$
there remains
\[
-\beta(\h x,\h y,\h z,p)+\alpha(\h x,\h y,\h z,p)/\alpha(x,y,z,p)+
c_1(x,y,z,p)/\alpha(x,y,z,p)+B'(\h x,\h y,\h z,x,y,z)=0.
\]
This equation is similar to \eqref{e-maineq-11} but has the additional
$B'$ term. Thus the solution is also a bit more general,
\[
\alpha(x,y,z,p)=\alpha_1(p)a(x,y,z),\quad
\beta(x,y,z,p)=\beta_1(p)+b(x,y,z),
\]
with suitable values for the auxiliary functions $c_1,B'$. Form
the coefficient of $\t z$ we first find that $\beta_1(p)$ is a constant
and then get the equation
\[
a(\h x,\h y,\h z)/\h z^2[\alpha_1(p)(g(p)+\h x)-\alpha_1(q)(g(q)+\h
x)]
=c_2(x,y,z,p)-c_2(x,y,z,q).
\]
There are now various way to solve this. \comment{see {\tt fthcc2}}

1: If $a(x,y,z)=z^2a_0$ then $\alpha_1(p)$ must be a constant, which
can be absorbed into $a_0$. We can take $g(p)\propto p$ and then we
get a solution with an extra term in comparison with
\eqref{eq:cc-r12}:
\begin{equation}
  \label{eq:cc-r41}
  \h{\t z}=-z \h {\t x}+z\frac{p\h z-q\t z}{\t z-\h z}.\tag{C-2.1}
\end{equation}
The triply shifted quantities are
\begin{subequations}
\begin{eqnarray}
\b{\h{\t x}}&=&\frac{
\t X(q\b z-r\h z)+\h X(r\t z-p\b z)+\b X(p\h z-q\t z)}
{\t X(\h z-\b z)+\h X(\t z-\h z)+\b X(\t z-\h z)},\\
\b{\h{\t y}}&=&\frac1z+\frac1z\frac{
\t z(q-r)+\h z(r-p)+\b z(p-q)}
{\t X(\h z-\b z)+\h X(\t z-\h z)+\b X(\t z-\h z)},\\
\b{\h{\t z}}&=&\frac{
\t X\h z\b z(q-r)+\h X\b z\t z(r-p)+\b X\t z\h z(p-q)}
{\t X(\h z-\b z)+\h X(\t z-\h z)+\b X(\t z-\h z)},
\end{eqnarray}
\end{subequations}
where we have used the shorthand notation $\t X=\ x-p,\, \h X=\h
x-q,\, \b X=\b x-r$.

2: If $a(x,y,z)=z^2a_0/x$ then we must have
$g(p)=\alpha_0/\alpha_1(p)$. After choosing $\alpha_1(p)\propto p$ and
scaling this yields the solution
\begin{equation}
  \label{eq:cc-r42}
  \h{\t z}
%=-d_0\frac{z}x+z\,\frac{p\h z-q\t z}{\t z-\h z}
%+\frac{z^2}x\,\frac{p\t y\h z-q\h y\t z}{\t z-\h z}
=-d_0\frac{z}x+
\frac{z}x\,\frac{p\t x\h z-q\h x\t z}{\t z-\h z}.\tag{C-2.2}
\end{equation}
If $d_0=0$ we can divide this equation with the $\h{\t x}$ equation
and write the result in terms of $w:=z/x$ in the form of lattice KdV.

For triply shifted quantities we get
\begin{subequations}
\begin{eqnarray}
  \b{\h{\t x}}&=&-d_0\frac{z}x \b{\h{\t y}}-
  \frac{\t x\h x\b z(p-q)+\h x\b x\t z(q-r)+\b x\t x\h z(r-p)}
  {p\t x(\h z-\b z)+q\h x(\t z-\h z)+r\b x(\t z-\h z)},\\
  \b{\h{\t y}}&=&\frac{x}z\,\frac
  {\t x(\h z-\b z)+\h x(\t z-\h z)+\b x(\t z-\h z)}
  {p\t x(\h z-\b z)+q\h x(\t z-\h z)+r\b x(\t z-\h z)},\\
\b{\h{\t z}}&=&-d_0\frac{
p\t z(\h z-\b z)+q\h z(\b z-\t z)+r\b z(\t z-\h z)}
  {p\t x(\h z-\b z)+q\h x(\t z-\h z)+r\b x(\t z-\h z)}\nonumber\\
&&-\frac{
p\t x\h z\b z(q-r)+q\h x\b z\t z(r-p)+r\b x\t z\h z(p-q)}
  {p\t x(\h z-\b z)+q\h x(\t z-\h z)+r\b x(\t z-\h z)}.
\end{eqnarray}
\end{subequations}

Both \eqref{eq:cc-r41} and \eqref{eq:cc-r42} provide a closed system
in $x,z$ so they are really two component systems.

3: If $a(x,y,z)$ does not have either of the other forms then both
$\alpha_1(p)$ and $g(p)$ must be constants. This leads to a trivial
solution: One finds first that $z\h{\t z}=-a(x,y,z)(\h{\t x}+g_1)$ but
then the third shift yields $\b{\h{\t x}}+g_1=0,\,\b{\h{\t z}}=0$.

\paragraph{Case C.3:} $\alpha(x,y,z,p)=a(p)z/y,\,
\gamma(x,y,z,p)=a(p)[g_1(p)x+g_2(p)]/y,\,g_1\neq 1$.  Since the
$x,y,z$-dependency is now explicit it is rather easy to solve the
remaining equations. By considering different powers of $\h y,\h
z$ one finds that $g_1(p)=d_1/a(p)$, that $\beta$ cannot depend on the
spectral parameter and can therefore be eliminated, and that
$g_2(p)=d_2/a(p)$. One also finds values for the auxiliary quantities
$c_1,c_2,c_3,B$. The resulting equation is \comment{see {\tt ajo.fthcc3}}
\begin{equation}
  \label{eq:c43-r}
  \h{\t z}=\frac{d_2x+d_1}{y}+
\frac{z}y\,\frac{p\t y\h z-q\h y\t z}{\t z-\h z}.\tag{C-3}
\end{equation}
If $d_i=0$ we recover the mBSQ/SBSQ equation \eqref{mBSQ-l}.
\comment{$d_2$ can be eliminated with a shift in spectra parameter.}
In this case we get a bit simpler results for the triply-shifted
quantities if we eliminate the shifted $x$-variables.
\begin{subequations}
\begin{eqnarray}
\b{\h{\t x}}&=&\frac{d_2x+d_1}{y}\b{\h{\t y}}+x+z\frac
{\t z\h y\b y(q-r)+\h z\b y\t y(r-p)+\b z\t y\h y(p-q)}
{\t z(q\h y-r\b y)+\h z(r\b y-p\t y)+\b z(p\t y-q\h y)},\\
\b{\h{\t y}}&=&y\frac{
\t z(\h y-\b y)+\h z(\b y-\t y)+\b z(\t y-\b y)}
{\t z(q\h y-r\b y)+\h z(r\b y-p\t y)+\b z(p\t y-q\h y)},\\
\b{\h{\t z}}&=&d_2z+(d_1+d_2 x)\frac{(q-r)\t z+(r-p)\h z+(p-q)\b z}
{\t z(q\h y-r\b y)+\h z(r\b y-p\t y)+\b z(p\t y-q\h y)}\\
&&+zy\frac{
qr\t z(\h y-\b y)+rp\h z(\b y-\t y)+pq\b z(\t y-\b y)}
{\t z(q\h y-r\b y)+\h z(r\b y-p\t y)+\b z(p\t y-q\h y)}.
\end{eqnarray}
\end{subequations}

\paragraph{Case C.4:} $\alpha(x,y,z,p)=a(p)[x+g_3(p)]z/y,\,
\gamma(x,y,z,p)=a(p)[g_2(p)+(x+g_3(p))(x+g_1(p))]/y,\,g_2\neq 0$. This
is similar to Case C.3. One first finds that $\beta=0$, then that
$a(p)$ is a constant, which can be scaled to $-1$,
$g_3(p)=d_0-g_1(p)$ and $g_2(p)=g_1(p)^2-d_0g_1(p)+d_1'$. Since
$p$-dependency enters only through $g_1(p)$ we put $g_1(p)=p+d_0/2$,
furthermore by $x$-translation we can eliminate $d_0$. The result is
therefore \comment{{\tt fthcc4} and ajo}
\begin{equation}
  \label{eq:c44-r}
  \h{\t z}=\frac{x\h{\t x}+ d_1}{y}+
\frac{z}y\,\frac{p\t y\h z-q\h y\t z}{\t z-\h z}.\tag{C-4}
\end{equation}
This can be written in a reversal symmetric form, if we replace the
$z$ on the RHS using $z=(\t x-\h x)/(\t y-\h y)$, which yields
\begin{equation}
  \label{eq:c44-ra}
  \h{\t z}y=x\h{\t x}+ d_1+
\frac{(\t x-\h x)(p\t y\h z-q\h y\t z)}{(\t y-\h y)(\t z-\h z)}.\tag{C-4r}
\end{equation}

Furthermore:
\begin{subequations}
\begin{eqnarray}
\b{\h{\t x}}&=&d_1\,\b{\h{\t y}}/y+
[
\t z (q (\h x-x) \b x-r (\b x-x) \h x)+
\h z (r (\b x-x) \t x-p (\t x-x) \b x)+\nonumber\\
&&\hskip 1.8cm \b z (p (\t x-x) \h x-q (\h x-x) \t x)
]/\Delta,\\
\b{\h{\t y}}&=&y [\t z (\h x-\b x)+\h z (\b x-\t x)+\b z (\t x-\h x)]/\Delta,\\
\b{\h{\t z}}&=&\{d_1 z [\t z (\h x+q-\b x-r)+\h z (\b x+r-\t x-p)+\b z (\t x+p-\h x-q)]\nonumber\\
&&\hskip 0.3cm+z [
\t z (q \h x (\b x+r)-r \b x (\h x+q))+
\h z (r \b x (\t x+p)-p \t x (\b x+r))+\nonumber\\&&
\hskip 0.9cm \b z (p \t x (\h x+q)-q \h x (\t x+p))
]
\}/\Delta,\\
\text{where}&&\Delta=
\t z [(q-x) (\h x-x)-(r-x) (\b x-x)]+\nonumber\\&&\hskip 0.9cm 
\h z [(r-x) (\b x-x)-(p-x) (\t x-x)]+\nonumber\\&&\hskip 0.9cm 
\b z [(p-x) (\t x-x)-(q-x) (\h x-x)].
\end{eqnarray}
\end{subequations}

\section{Discussion}
We have considered multi-dimensionally consistent equations generated
by quasilinear equations of three variables defined on the edges of
the consistency cube. Assuming constant coefficients and dependency on
only two of the three possible shifted variables these equations were
first reduced into three canonical cases \eqref{e-11}, \eqref{e-c5},
or \eqref{e-12fin}. In each case, starting with the equations given on
the 12 edges of the consistency cube, we can propagate $x$ to other 7
corners of the cube, which leaves 5 equations. Assuming free initial
values for $y,\,\t y,\,\h y,\,\b y$ we can use 4 of the remaining
equations to obtain $\h{\t y},\, \b{\h y},\, \b{\t y},\, \b{\h{\t y}}$
and the final equation gives the first consistency requirement for
$z$, \eqref{e-c1.1-cac}, \eqref{e-c5-cac}, or \eqref{e-c1.2-cac},
respectively.  The equations for the shifted $z$'s are equations on
the sides of the cube with additional consistency conditions, this
implies equations for the remaining freedom in the $F$ function, given
in \eqref{e-c1.1-zcac}, \eqref{e-c5-zcac}, \eqref{e-c1.2-zcac},
respectively. These functional equations were solved with the results
given above. The results can be summarized as follows:

\paragraph{One component results.}
In three cases the resulting $z$ equation only involved $z$-variables,
see \eqref{sol-A-1} (KdV), \eqref{eq:cc-r11}, and \eqref{eq:cc-r12}
(mKdV). For these equations one can reverse the point of view and
state that for these $z$-equations one can adjoin edge equations (of
the canonical form) consistently.  One may ask whether any consistent
1-component equation (such as those in the ABS list \cite{CAC3}) can be so
extended, if more freedom is given for the form of the edge equations.

\paragraph{Quasilinear z-equation.} In the case of \eqref{s-B-1}, or
its two canonical forms \eqref{s-B-1acan} and  \eqref{s-B-1bcan}, the 
$z$ equations are also quasilinear, and in fact reversal symmetric.
However, these equations do not allow evolution towards lower right or
upper left, because then these equations do in fact impose further
conditions on the initial values.  The triply shifted quantities
\eqref{eq:cac-B-1a} or \eqref{eq:cac-B-1b} suggest periodic reduction
along the diagonal.

\paragraph{Two component results.}  For Case C we found two essentially
2-component results
\begin{equation}
  \label{eq:cc-r41-b}
  \h{\t z}=-z \h {\t x}+z\frac{p\h z-q\t z}{\t z-\h z},\tag{C-2.1}
\end{equation}
and 
\begin{equation}
  \label{eq:cc-r42b}
  \h{\t z}=-d_0\frac{z}x+
\frac{z}x\,\frac{p\t x\h z-q\h x\t z}{\t z-\h z},\tag{C-2.2}
\end{equation}
since together with the $x$-equation in \eqref{e-c1.2-pq} the
variables $x,z$ form a closed system. However, since neither $\h {\t
  x}$ nor \eqref{eq:cc-r41-b} involves $x$, this system does not
define evolution in all directions. If $d_0=0$ \eqref{eq:cc-r42b} can
be reduced to lmKdV for $w:=z/x$.

\paragraph{Fully three component results.}
We found four fully three-component systems that allow evolution in
all directions. One criterion for this is that the $z$ equation must
involve un-shifted $y$, which does not appear in any of the edge
equations.

$\bullet$ For the base-equation \eqref{e-11}, i.e., $\t x z=\t
y+x$, we found the
$z$-equation, 
\begin{equation}\label{sol-A-3-b}
%\boxed
{\h{\t z}=\frac{y}x+\frac1x\,\frac{p\,\t x-q\,\h x}{\t z-\h z},}
\tag{A-2}
\end{equation}
or in the reversal symmetric form \eqref{sol-A-3r}.

$\bullet$ For the base equation \eqref{e-c5}, i.e., $x\t x=\t y+z$, we
found a mild generalization of the lBSQ equation \eqref{lBSQ-l}:
\begin{equation}\label{s-B-2-2}
\h{\t z}+y=b_0(\h{\t x}-x) +x\,\h{\t x}+\frac{p-q}{\t x-\h x},\tag{B-2}
\end{equation}

$\bullet$ For the C-case, i.e., $z\t y=\t x-x$, we found two
equations, which are kind of modifications of the lmBSQ/lSBSQ equation
\eqref{mBSQ-l}:
\begin{equation}
  \label{eq:c43-r2}
  \h{\t z}=\frac{d_2x+d_1}{y}+
\frac{z}y\,\frac{p\t y\h z-q\h y\t z}{\t z-\h z},\tag{C-3}
\end{equation}
and
\begin{equation}
  \label{eq:c44-r2}
  \h{\t z}=\frac{x\h{\t x}+ d_1}{y}+
\frac{z}y\,\frac{p\t y\h z-q\h y\t z}{\t z-\h z}.\tag{C-4}
\end{equation}

\section*{Acknowledgments} 
I would like to thank K. Kajiwara, F. Nijhoff, and C. Viallet for discussions.

{\small
\appendix

\section{Classification of the quasi-linear part}
Introducing vector notation
\[
\t u=(\t x,\t y,1)^t,\quad U=(x,y,z,1)^t,
\]
we can write the equation as
\begin{equation}
(\t u)^t\, A\, U=0, \text{ where }
%(\h x,\t y,1)
A=\begin{pmatrix}
a_{11} & a_{12} & a_{13} & b_{1}\\
a_{21} & a_{22} & a_{23} & b_{2}\\
c_{1} & c_{2} & c_{3} & d
\end{pmatrix}.
%\begin{pmatrix}x\\y\\z\\1\end{pmatrix}=0.
\end{equation}

The rotation freedom means that we can introduce primed quantities by
\begin{equation}
  \label{e-tra1}
\t u=m\,\t u',\quad U=M\,U',\text{ where }
%  \begin{pmatrix}\t x'\\ \t y'\\ 1\end{pmatrix}=
m=\begin{pmatrix}
s_{11} & s_{12} & t_x \\
s_{21} & s_{22} & t_y \\
0 & 0 & 1 \end{pmatrix},
%  \begin{pmatrix}\t x\\ \t y\\ 1\end{pmatrix},\quad
%  \begin{pmatrix} x'\\ y'\\ z'\\  1\end{pmatrix}=
\, M=
\begin{pmatrix}
s_{11} & s_{12} & 0 & t_x \\
s_{21} & s_{22} & 0 & t_y \\
o_1 & o_2 & o_3 & t_z \\
0 & 0 & 0 & 1 
\end{pmatrix}
%  \begin{pmatrix}x\\ y\\ z\\ 1\end{pmatrix}
\end{equation}
and this induces the transformation $A\to A'=m^t\,A\,M$. 

\subsection{Classification of the quadratic part}
Let us first consider the quadratic part, which is determined by
\begin{equation}
A_q=\begin{pmatrix}
a_{11} & a_{12} & a_{13}\\
a_{21} & a_{22} & a_{23}
\end{pmatrix},
\end{equation}
and which can be transformed as  $A_q\to A_q'=m_q^t\,A_q\,M_q$ by
\begin{equation}
  \label{e-tra1b}
m_q=\begin{pmatrix}
s_{11} & s_{12}\\
s_{21} & s_{22}
\end{pmatrix},
\, M_q=
\begin{pmatrix}
s_{11} & s_{12} & 0\\
s_{21} & s_{22} & 0\\
o_1 & o_2 & o_3
\end{pmatrix}.
\end{equation}

The main classification is as follows: \comment{file {\tt rot}}

\noindent $\bullet$ If one or both of
$a_{13},\,a_{23}$ are nonzero then we can choose $s_{ij}$ so
that $a_{13}s_{11}+a_{23}s_{21}=0$ while
$a_{13}s_{12}+a_{23}s_{22}\neq 0$ and $s_{11}s_{22}-s_{12}s_{21}\neq
0$. Then using $o_1,o_2,o_3$ we can put $A_q'$ into the form
\[
A_q=\begin{pmatrix}
f & g & 0\\
0 & 0 & 1
\end{pmatrix},
\]
This we divide into three cases:

{\bf Case 1:} $f=g=0$ after which we exchange $x,y$ and get
\[
\t x z= \text{(linear terms)}.
\]

{\bf Case 2:}  $f=0, g\neq 0$
\[
\t x y+\t y z=  \text{(linear terms)}.
\]

{\bf Case 3:} Finally if  $f\neq 0$ we can use the freedom
still left in $s_{ij}$ to put $g=0$:
\[
\t x x +\t y z=  \text{(linear terms)}.
\]

\noindent $\bullet$ Next if $a_{13}=a_{23}=0$ then $o_1,o_2$ are
superfluous and we only have the $s$-part to operate with. Let us
denote the left $2\times2$ block of $A_q$ by $A_s$, then the remaining
transformation is $s^t A_s s$. This can be used to bring $A_s$ into
one of the following 6 normal forms \comment{proof in {\tt sqa}}
\[
\begin{pmatrix} 0 & 0 \\ 0 & 0 \end{pmatrix},
\begin{pmatrix} 0 & 1 \\ 0 & 0 \end{pmatrix},
\begin{pmatrix} 1 & 0 \\ 0 & 0 \end{pmatrix},
\begin{pmatrix} 1 & 0 \\ 0 & 1 \end{pmatrix},
\begin{pmatrix} 1 & \alpha \\ 0 & 1 \end{pmatrix},
\begin{pmatrix} 0 & 1 \\ -1 & 0 \end{pmatrix},
\]
where the parameter $\alpha$ is nonzero. The first case must be
omitted, as it leads to linear equation. Thus we have 5 more cases:

{\bf Case 4:}
\[
\t x y= \text{(linear terms)}.
\]

{\bf Case 5:}
\[
\t x x= \text{(linear terms)}.
\]

{\bf Case 6:}
\[
\t x x+\t y y= \text{(linear terms)}.
\]

{\bf Case 7:}
\[
\t x x+\alpha \t x y+\t y y= \text{(linear terms)}.
\]

{\bf Case 8:}
\[
\t x y-\t y x= \text{(linear terms)}.
\]

\subsection{Detailed classification}
We will next do a detailed analysis and classification of the
quasilinear equations. \comment{ REDUCE files used in this section:  {\tt
  caf, moe}}

\subsubsection*{Case 1}
In Case 1 we must have $b_2\neq0$ in order to have $\t y$ dependency,
therefore we scale to $b_2=-1$. This case is invariant under the full
transformation \eqref{e-tra1b}, and if we take $s_{11}=1,\,
s_{12}=0,\,s_{22}=1,\,o_1=o_2=0,\,o_3=1,\, t_z=s_{21}-b_1,\,t_x=-c_3$
we can simplify the linear terms into
\begin{equation}\label{case1yl}
\t xz-\t y+(Bt+A)x+By+s(B-1)+C=0.
\end{equation}
This is further divided as follows:

If $B=0$ $y$ disappears and we can eliminate the
constant using $s$. Now we must have $A\neq 0$ otherwise the system
leads to triviality, thus we can scale to
\begin{equation}
  \label{e-11dup}
\t x z=\t y+x,
\end{equation}
which is Case A

If $B\neq0$ we can simplify the system by eliminating $x$ using
$t$ and obtain
\begin{equation}
  \label{e-12apu}
 \t x z=\t y-By-s(B-1)-C.
\end{equation}
If $B\neq 1$ we can next eliminate the constant $C$ with $s$ and after
that redefine $(x,y)\mapsto (B^{n+m+k}y, B^{n+m+k}x)$ obtaining
\begin{equation}
  \label{e-12findup}
 \t y z=\t x-x,
\end{equation}
which is Case C.  If instead $B=1$ we can eliminate $C$ with
$(x,y)\mapsto (y+(n+m+k)C,x)$, which again yields \eqref{e-12fin}.

\subsubsection*{Case 2}
The quadratic part of Case 2 is invariant under \eqref{e-tra1b} if
$s_{11}=1,\, s_{21}=0,\, s_{22}=1$. Choosing furthermore
$s_{12}=-o_2,\,o_1=0,\, o_3=1,\, t_z=b_2,\,t_y=-b_1,\, t_x=o_2(b_1 + c_1 - c_3)
- c_2$
the equation becomes
\[
  \t x\, y+\t y\,z=A'x+B'z+o_2(B'-A')A'+C'.
\]

If $B'\neq 0$ the transformation \comment{ see {\tt moe}}
\[
(x,y,z)\mapsto\left(\frac{xt+y+s}{x+1},\frac{B'}{x+1},
\frac{-A'/B'(xt+y+s)-D'(x+1)+z+t}{x+1}\right)
\]
takes this to \eqref{case1yl} with $B=A'/B',A=C=D'$ where 
$D'=o_2A'(B'-A')/B'+C'/B'$.

If $B'=0$, $A'\neq 0$ then we can eliminate the constant term with $o_2$
and scale to
\begin{equation*}
%  \label{e-21b}
  \t x\, y+\t y\,z=x.
\end{equation*}
This can be transformed to \eqref{e-11} by $(x,y,z)\mapsto (1/y,-x/y,-z/y)$.

If $A'=B'=0$ then $C'\neq 0$ and can be scaled to 1 and we have
\begin{equation*}
%  \label{e-21a}
  \t x\, y+\t y\,z=1.
\end{equation*}
With $(x,y,z)\mapsto(y/x,1/x,z/x)$ this can be transformed to the
third canonical form
\begin{equation}
  \label{e-c5dup}
 \t x\,x=\t y+z,
\end{equation}
which is Case B.

\subsubsection*{Case 3}
For Case 3 and the other remaining cases the possible rotational
freedom in $m_q$ is not useful. Using the translational freedom with
$o_3=1,t_z=-b_2,t_x=-b_1,t_y=-c_3$ we can bring the equation into the
form
\begin{equation}
  \label{e-c3}
  \t x\,x+\t y\,z=Ax+By+C,
\end{equation}

If $C=0$ but $AB\neq 0$ then the equation is taken to \eqref{e-11}
with the transformation 
\[ 
(x,y,z)\mapsto( -A(x+1)/(y-1),\,(A^2/B)x/(y-1), -B(x+z+1)/(y-1).
\]

If $C=B=0$ the system is trivializes with $(x,y,z)\mapsto (x+A,y,z(x+A))$.

If $C=A=0$ we scale to $B=1$ and get
\begin{equation}
  \label{e-c31-a}
\t x\,x+\t y\,z=y.
\end{equation}
This is transformed to \eqref{e-c5} with $(x,y,z)\mapsto(x/y,1/y,-z/y)$:

Now we assume $C\neq0$ and scale it to 1:
\begin{equation}
  \label{e-c32a}
  \t x\,x+\t y\,z=Ax+By+1,
\end{equation}
Here, for convenience, we parametrize $A=-F+1/F$, then we can
transform \eqref{e-c32a} to \eqref{e-12apu} (with $B=-F^2,s=0,C=-1$) using
\[
(x,y,z)\mapsto\left(
\frac{-(F^2y+1)}{Bx+Fy},
\frac{-x}{Bx+Fy},
\frac{-BF^3y+z-B(F-1/F)}{Bx+Fy},
\right).
\]
%The subcase $A=0$ is included as $F=1$.

\subsubsection*{Case 4}
In this case we must have $b_2\neq0$. We use the freedom in the $z$
transform $o_i$ and the translations $t_i$ to bring the equation into
\[
  \t x\,y=\t y+(Ao_1+B)x+Az+(At_z+C).
\]
Now if $A=0$ the variable $z$ disappears completely, which is not
allowed, thus using $o_1,t_z$ and scaling we obtain
\begin{equation*}
%  \label{e-c4}
  \t x\,y=\t y+z,
\end{equation*}
and with $(x,y,z)\mapsto( -y/x, 1/x,-z/x)$ this becomes \eqref{e-11}.

\subsubsection*{Case 5}
Here we must have $b_2c_3o_3\neq 0$. We can then choose
$o_1,o_2,t_z,t_x$ and scaling so that we obtain  \eqref{e-c5}.

\subsubsection*{Case 6}
The situation is similar to Case 5 and leads first to
\begin{equation}
  \label{e-c6}
  \t x\,x+\t y y=z.
\end{equation}
With $(x,y,z)\mapsto((y+1)/x, iy/x,\, z/x)$ this gives the  $B=-1,C=-1$
subcase of \eqref{e-12apu}.

\subsubsection*{Case 7}
Now using the translation freedom and
$s_{11}=0,s_{12}=-1,s_{21}=1,s_{22}=\alpha$ we get
\begin{equation}
  \label{e-c7}
  \t x\,x+\alpha\t x y+\t y y=z,\quad \alpha\neq 0.
\end{equation}
For convenience let us parametrize $\alpha=i(\kappa-1/\kappa)$. Then
with transformation $(x,y,z)\mapsto((\kappa y+1)/x,iy/x,\kappa z/x)$ this
case becomes \eqref{e-12apu} with $B=-\kappa^{-2},\,C=-\kappa^{-1}$.

\subsubsection*{Case 8}
The rotational freedom is not useful but we can effectively use
translations and get
\begin{equation}
  \label{e-c8}
  \t x\,y-\t y x=z.
\end{equation}
With $(x,y,z)\mapsto (1/y, -x/y, z/y)$ this becomes \eqref{e-12fin}.

In conclusion, by suitable transformations we can bring all nontrivial
models satisfying the condition mentioned at the beginning of Sec.~2
to one of the 4 forms \eqref{e-11}, \eqref{e-12fin}, or \eqref{e-c5}.

\section{Solving functional equations}

\subsection{$\h\alpha+\alpha \h \beta=c$\label{sol11}}
This is the equation needed in Case A, 
\begin{equation}
  \label{e-def-11}
  \alpha(\h x,\h x z-x,\h z,p)+\alpha(x,y,z,p)\,\beta(\h x,\h x z-x,\h
  z,p) =c_1(x,y,z,p).
\end{equation}
It is obvious that this has the solution $\alpha\equiv 0$.
Furthermore, since 
\begin{equation}\label{dep-lemma}
f(\h x,\h x z-x,\h z,p)=g(x,y,z,p) \Rightarrow f=f(p),
\end{equation}
we also find that if $\alpha\not\equiv 0$ then
$\alpha(x,y,z,p)=\alpha(p) \Leftrightarrow \beta(x,y,z,p)=\beta(p)$
and this provides another solution.

Next let us assume that both $\alpha$ and $\beta$ depend non-trivially
on some of the other possible variables. Computing the $y$-derivative
of \eqref{e-def-11} shows that since $\beta$ must depend on some of
its first three variables $\alpha$ cannot depend on its second
variable. Thus we get 
\begin{equation}
  \label{e-s1-11}
  \alpha(\h x,\h z,p)+\alpha(x,z,p)\,\beta(\h x,\h x z-x,\h z,p)  
  =c_2(x,z,p)
\end{equation}
Now by operating on this with $(\h x\partial_x+\partial_z)\partial_{\h
  z}$ we obtain
\[
\left[\h x\partial_x\alpha(x,z,p)+\partial_z\alpha(x,z,p)\right]
\partial_{\h z}\beta(\h x,\h x z-x,\h z,p) =0
\]
If the first factor vanishes then we are back to $\alpha=\alpha(p)$
and therefore we must assume that $\beta(x,y,z,p)=\beta(x,y,p)$. But
then it follows that $\alpha(x,z,p)=\alpha(x,p)$ and the equation
reduces to
\begin{equation}
  \label{e-s2-11}
  \alpha(\h x,p)+\alpha(x,p)\,\beta(\h x,\h x z-x,p)  =c_2(x,z,p)
\end{equation}
Now computing the $z$-derivative and then applying \eqref{dep-lemma}
we find that $\beta(x,y,z,p)=(y/x)b_1(p)+b_0(x,p)$ and then the
equation becomes, after some rearrangements,
\begin{equation}
  \label{e-s3-11}
  \alpha(\h x,p)+\alpha(x,p)[-(x/\h x)b_1(p)+b_0(\h x,p)]  
=c_2'(x,z,p)
\end{equation}
Now it is clear that  
$\alpha(x,y,z,p)=a_1(p)/x+a_0(p)$ and $b_0(\h x,p)=b_2(p)/\h x+b_3(p)$.
Next we find that since $a_1\neq 0$ we must have $b_2=b_0=0,\,b_1=1$.

Thus collecting the results we find that equation \eqref{e-def-11} has
precisely the following solutions:
\begin{itemize}
\item $\alpha(x,y,z,p)\equiv0$,
\item $\alpha(x,y,z,p)=\alpha(p),\,\beta(x,y,z,p)=\beta(p)$,
\item $\alpha(x,y,z,p)=\alpha(p)/x,\,\beta(x,y,z,p)=y/x+\beta(p)$.
\end{itemize}

\subsection{$\h\alpha+\beta\h\beta=c$\label{sol5}}
The equation to solve for Case B is
\begin{equation}\label{e-s2-def}
\alpha(\h x, x \h x-z,\h z,p)+\beta(x,y,z,p)\beta(\h x, x \h x-z,\h z,p)
=c_3(x,y,z,p).
\end{equation}
Again we have the solutions $\alpha=\alpha(p),\, \beta=\beta(p)$ and
next we assume that there is some other dependency as well.

Computing the $y$-derivative and recalling \eqref{dep-lemma} we find
that $\beta$ cannot depend on its second variable  and the equation
becomes
\begin{equation}\label{e-s2-2}
\alpha(\h x, x \h x-z,\h z,p)+\beta(x,z,p)\beta(\h x,\h z,p)
=c_3(x,y,z,p).
\end{equation}

Now dividing both sides with $\beta(x,z,p)$ and then operating with
$(\partial_x+\h x\partial_z)\partial_{\h z}$ we get
\begin{equation}\label{e-s2-3}
0=
\partial_{\h z}\alpha(\h x, x \h x-z,\h z,p)
\left[(\partial_x+\h x\partial_z)1/\beta(x,z,p)\right].
\end{equation}
Since $\beta$ cannot just depend on $p$ the first factor must vanish
and thus $\alpha$ cannot depend on $\h z$, then from \eqref{e-s2-2} we
conclude that neither can $\beta$. Thus we find
\begin{equation}\label{e-s2-4}
\alpha(\h x, x \h x-z,p)+\beta(x,p)\beta(\h x,p)
=c_3(x,y,z,p).
\end{equation}
Next computing the $z$-derivative of this we find that 
$\alpha(x,y,z,p)=ya_1(p)+a_2(x,p)$, thus we get
\begin{equation}\label{e-s2-5}
x\h xa_1(p)+a_2(\h x,p)+\beta(x,p)\beta(\h x,p)
=c_3'(x,y,z,p).
\end{equation}
Since $\beta$ has to depend on its first variable this dependency must
be linear, $\beta(x,y,z,p)=xb_1(p)+b_0(p)$, leading to
\begin{equation}\label{e-s2-5x}
x\h xa_1(p)+a_2(\h x,p)+[xb_1(p)+b_0(p)][\h xb_1(p)+b_0(p)]
=c_3'(x,y,z,p).
\end{equation}
Now also $a_2(\h x,p)$ has to be linear and the result is therefore
\begin{equation}\label{s5-res}
\left\{\begin{array}{rcl}
\alpha(x,y,z,p)&=&a_0(p)-xb_0(p)b_1(p)-yb_1(p)^2\\
\beta(x,y,z,p)&=&b_0(p)+xb_1(p)\end{array}\right.
\end{equation}

With the special case $b_1(p)=0$ we get the solution mentioned at
the beginning of this section.

\subsection{$\h\alpha (z/\h z)(\h y\alpha+\gamma)-\h \gamma \alpha=
c_3\alpha-c_1\gamma$\label{sol-cc}}
The equation to solve for Case C is \comment{tarkista oikean puolen merkit}
\begin{align}
  \alpha(\h y z+x,\h y,\h z,p)\,(z/\h z)\,[\alpha(x,y,z,p)\h
  y+\gamma(x,y,z,p)]
  -\gamma(\h y z+x,\h y,\h z,p)\,\alpha(x,y,&z,p)=\nonumber\\
  -\alpha(x,y,z,p) c_3(x,y,z,p)+\gamma(x,y,z,p)c_1(x,y,z,p).&
  \label{eq:cc-eqd}
\end{align}
Clearly this has the solution $\alpha\equiv0$. Next we assume
$\alpha\neq 0$.  After operating on \eqref{eq:cc-eqd} divided by
$\alpha(x,y,z,p)$ by $\partial_y\partial_{\h z}$ the result factorizes
as \comment{see {\tt solcc32x}}
\begin{align}
  \label{eq:s-cc-1}
  [\alpha(x,y,z,p)\,\partial_y \gamma(x,y,z,p)-
  \gamma(x,y,z,p)\,\partial_y \alpha(x,y,z,p)]\times\nonumber&\\
  [\alpha(\h y z+x,\h y,\h z,p)-\h z\,\partial_{\h z} \alpha(z\h
  y+x,\h y,\h z,p)]&=0
\end{align}

a) Let us first consider the first factor and assume that the second
factor does not vanish. Then clearly \comment{see {\tt solcc31}}
\begin{equation}
  \label{eq:cc-s3b}
\gamma(x,y,z,p)=\alpha(x,y,z,p)g(x,z,p).  
\end{equation}
Now operating on \eqref{eq:cc-eqd} divided by $\alpha(x,y,z,p)$ by
$(\partial_z-\h y\partial_x)\partial_{\h z}$ the result factorizes as
\begin{align}
  \label{eq:s-cc-2}
  [z\h y\partial_x g(x,z,p)-z\partial_z g(x,z,p)-g(x,z,p)-\h y]\times
  \nonumber&\\
  [\alpha(\h y z+x,\h y,\h z,p)-\h z\partial_{\h z}\alpha(\h y z+x,\h y,\h
  z,p)]&=0
\end{align}
Again we use the first factor, and since $g$ does not depend on $\h y$
we get two equations
\[
z\partial_z g(x,z,p)+g(x,z,p)=0,\quad z\partial_x g(x,z,p)-1=0.
\]
which are solved by 
\begin{equation}
  \label{eq:cc-s4}
g(x,z,p)=(x+g(p))/z.  
\end{equation}
This yields the solution
\begin{equation}
  \label{eq:cc-r1}
\alpha=\text{arbitrary},\quad \gamma(x,y,z,p)=\alpha(x,y,z,p)(x+g(p))/z.
\end{equation}

b) Next we take the second factor in \eqref{eq:s-cc-1}, which is
solved by \comment{see{\tt solcc32x}}
\begin{equation}
  \label{eq:cc-2-1a}
\alpha(x,y,z,p)=z\,a(x,y,p).  
\end{equation}
Then operating by $\partial_{\h z}$ on \eqref{eq:cc-eqd} yields
\begin{equation}
  \label{eq:cc-2-1b}
a(x,y,p)\partial_{\h z} g(\h y z+x,\h y,\h z,p)=0,
\end{equation}
so that
\begin{equation}
  \label{eq:cc-2-1c}
\gamma(x,y,z,p)=g(x,y,p).  
\end{equation}
The $\h z$ derivative of \eqref{eq:cc-eqd} now vanishes and next we
consider its $\h y$ derivative, which yields
\begin{align}
  \label{eq:cc2-2}
\partial_{\h y}a(\h y z+x,\h y,p)[z\h y\,a(x,y,p)+g(x,y,p)]&
\nonumber\\
-\partial_{\h y}g(\h y z+x,\h y,p)a(x,y,p)+
z\,a(z\h y+x,&\h y,p)a(x,y,p)=0.
\end{align}
Since $z$-dependency is now explicit it is useful to change variables
by replacing $z$ with $\h x$ using $\h y z=\h x-x$, which also implies
$\partial_{\h y}\to\partial_{\h y}+z\partial_{\h x}$.  Equation
\eqref{eq:cc2-2} then becomes
\begin{align}
  \label{eq:cc2-3}
  [\h y\partial_{\h y}a(\h x,\h y,p)+(\h x-x)\partial_{\h x}a(\h y,\h
  x,p)] [\h x-x+g(x,y,p)/a(x,y,p)]&
  \nonumber\\
  -[\h y\partial_{\h y}g(\h x,\h y,p)+(\h x-x)\partial_{\h x}g(\h y,\h
  x,p)]+ (\h x-x)a(\h y,&\h x,p)=0.
\end{align}
Now computing the $y$-derivative of this yields
\begin{equation}
  \label{eq:cc2-4}
  [\h y\partial_{\h y}a(\h x,\h y,p)+(\h x-x)\partial_{\h x}a(\h
  y,\h x,p)] 
  [a(x,y,p)\partial_y g(x,y,p)-   g(x,y,p)\partial_y a(x,y,p)]=0.
\end{equation}
The first factor vanishes only if $a(x,y,p)=a(p)$ and then
\eqref{eq:cc2-3}  implies 
\begin{align}
  \label{eq:cc2-5}
  \h y\partial_{\h y}g(\h x,\h y,p)+\h x[\partial_{\h x}g(\h y,\h
  x,p)- a(p)]=0,\quad
  \partial_{\h x}g(\h x,\h y,p)-a(p)=0,
\end{align}
which is solved by
\[
g(x,y,p)=a(p)(x+g(p)).
\]
This yields, however, only a subcase of \eqref{eq:cc-r1}.

For the second factor in \eqref{eq:cc2-4} to vanish we take 
\begin{equation}
  \label{eq:cc-s3a}
g(x,y,p)=a(x,y,p)g_1(x,p).
\end{equation}
Substituting this into \eqref{eq:cc2-3} and 
then operating with $\partial_x^2$ yields
\begin{equation}
  \label{eq:cc-6}
  \h y\partial_{\h y}a(\h x,\h y,p)\partial_{x}^2 g(x,p)
  -\partial_{\h x}a(\h x,\h y,p)[
(x-\h x)\partial_{x}^2 g(x,p)+2\partial_{x} g(x,p)-2]=0
\end{equation}
We divide solving this into four subcases:

1. Let $\partial_{\h x}a(\h x,\h y,p)=\partial_{\h y}a(\h y,\h
x,p)=0$, i.e., $a=a(p)$. Then \eqref{eq:cc2-3} implies $g(x,p)=x+g(p)$, 
which again yields a subcase of \eqref{eq:cc-r1}.

2. Let $\partial_{\h x}a(\h x,\h y,p)=\partial_x^2g(x,p)=0$, i.e., 
$a(x,y,p)=a(y,p),\,g(x,p)=g_1(p)x+g_0(p)$. Equation \eqref{eq:cc2-3}
then reduces to $(g_1(p)-1)(\h y\partial_{\h x}a(\h y,p)+a(\h
y,p))=0$. This first factor again yields a subcase of
\eqref{eq:cc-r1} but the second factor yields a new solution
\begin{equation}
  \label{eq:cc-r2}
\alpha=a(p)z/y,\quad \gamma=a(p)(g_1(p)x+g_2(p))/y,\quad g_1(p)\neq 1.
\end{equation}

3. Next assume  $\partial_{\h x}a(\h x,\h y,p)\neq 0,
\partial_x^2g(x,p)=0$. Then we must in fact have $\partial_xg(x,p)=1$,
leading back to \eqref{eq:cc-r1}.

4. Finally assume  $\partial_{\h x}a(\h x,\h y,p)\neq 0,
\partial_x^2g(x,p)\neq0$. Then dividing \eqref{eq:cc-6} by their
product and computing a $x$ derivative yields the equation
\begin{equation}
  \label{eq:cc-7}
  2(\partial_x g-1)\partial_x^3g=3(\partial_x^2 g)^2.
\end{equation}
This has the solution
\begin{equation}
  \label{eq:cc-8}
  g(x,p)=x+g_1(p)+g_2(p)/(x+g_3(p)).
\end{equation}
Here we may assume $g_2\neq 0$, else we are back to a subcase of
\eqref{eq:cc-s4}. Returning to \eqref{eq:cc2-3} we get from the
coefficient of $x$ two equations,
\[
\h y\partial_{\h y}a(\h x,\h y,p)+a(\h x,\h y,p)=0,\quad
(\h x+g_3)\partial_{\h x}a(\h x,\h y,p)-a(\h x,\h y,p)=0,\quad
\]
which are solved by
\[
a(x,y,p)=a(p)(x+g_3(p))/y.
\]
This leads to the solution
\begin{equation}
  \label{eq:cc-r4}
  \alpha(x,y,z,p)=a(p)(x+g_3(p))z/y,\quad 
  \gamma(x,y,z,p)=a(p)[g_2(p)+(x+g_3(p))(x+g_1(p))]/y.
\end{equation}
} Thus we have obtained four solutions, $\alpha\equiv 0$,
\eqref{eq:cc-r1}, \eqref{eq:cc-r2}, and \eqref{eq:cc-r4}.

\end{document}